\documentclass[]{spie}  
\UseRawInputEncoding

\usepackage{amsmath,amsfonts,amssymb}
\usepackage{graphicx}
\usepackage[colorlinks=true, allcolors=blue]{hyperref}

\title{Dynamics and Fusion of Majorana Zero Modes in Quantum Dot-based Interacting Kitaev Chains.}
\author[a,b]{Bradraj Pandey}
\author[a,b]{Elbio Dagotto}
\affil[a]{Department of Physics and Astronomy, The University of Tennessee, Knoxville, Tennessee 37996, USA,}
\affil[b]{Materials Science and Technology Division, Oak Ridge National Laboratory, Oak Ridge, Tennessee 37831, USA}
\authorinfo{bradraj.pandey@gmail.com\\ edagotto@utk.edu}
\begin{document} 
\maketitle

\begin{abstract}
     Motivated by the recent experimental realization of a                              
      minimal Kitaev chain in quantum dot systems, we present our theoretical findings on the dynamics and                                        
      fusion of MZMs at or near the $``$sweet spot" $t_h = \Delta$ (where the fermionic hopping $t_h$ and superconducting coupling $\Delta$ are equal).
     We investigated the dynamics and fusion of MZMs using time-dependent real-space local density-of-states methods.                
     The movement of Majoranas and the detection of fusion channels are crucial for topological quantum computations.                
      Additionally, we discuss our recent discovery of exotic $``$multi-site" MZMs in $Y$-shaped Kitaev wires,                           
      which is important for the potential braiding of Majoranas in $Y$-junctions formed from arrays of quantum dots.                 
      Finally, we present results on "non-trivial" fusion using canonical Kitaev wires at the sweet spot.       
     \end{abstract}                                                                                                                    

\section{INTRODUCTION}
Topological superconductors (TSC) offer an ideal platform to realize exotic Majorana zero modes. 
These Majorana zero modes (MZMs) follow non-Abelian exchange statistics and have potential applications 
to realize fault-tolerant quantum computing~\cite{Kitaev1,Kitaev2,brien,Sarma,Nayak}. 
The most explored experimental method for creating Majorana zero modes is 
through hybrid semiconductor nanowire systems~\cite{Roman,Mourik}. 
In these systems, the MZMs are expected to appear at the edge of the nanowires by manipulating 
the external magnetic field in the presence of large spin-orbit coupling and in proximity to conventional 
$s$-wave superconductors~\cite{Mourik}. However, due to the presence of disorder and impurities, the 
realization of actual Majorana zero modes is still challenging~\cite{Stanescu,Gomanko}. 
The trivial near-zero energy states induced by disorder pose a serious hurdle in 
detecting the actual MZMs in the zero-bias conductance peak experiments~\cite{Gomanko,Loo,PanH}. 
Recently, an alternative approach to realize MZMs using a chain of quantum dots has been proposed~\cite{Jay,Loss,Deng}.
The minimal Kitaev chain using just two quantum dots, coupled with a short superconductor-semiconductor hybrid, 
has been successfully realized~\cite{Dvir}. Remarkably, in this experiment, 
a pair of Majorana zero modes were observed in the tunneling conductance measurements 
at the $``$sweet spot"  $t_h=\Delta$ 
(where the electronic hopping $t_h$ and superconducting coupling $\Delta$ are equal in magnitude)~\cite{Dvir}. 
Very recently, using three quantum dots, MZMs were realized, and it was found that the topological protection 
increased compared to the two-site Kitaev chain~\cite{Bordin,ChunX}. Recent progress in quantum dots, 
which can be precisely controlled using local gates, presents a promising avenue for quantum information 
processing using MZMs~\cite{Mazur,Liu}. 
Local control of individual quantum dots significantly reduced the effect of disorder and 
the detection of MZMs~\cite{Dvir,ChunX,Liu}.

  \begin{figure} [ht]                                                                                                            
     \begin{center}                                                                                                                 
    \begin{tabular}{c} 
      \includegraphics[height=5cm]{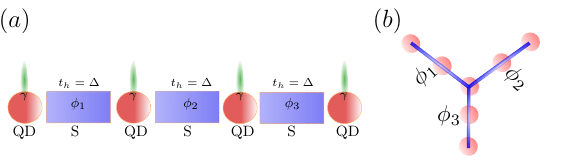}                                                                                        
      \end{tabular}                                                                                                                  
      \end{center}                                                                                                                   
\caption{(a) Schematic representation of a coupled quantum-dot system designed to realize Majorana zero modes ($\gamma$). The quantum dots (QD) are connected to a short superconductor (S) with phase ($\phi_i$), enabling effective tuning of the hopping parameter $t_h$ and the superconducting pairing term $\Delta$ between the quantum dots. (b) Illustration of the $Y$-shaped geometry utilizing seven quantum dots with different phases $\phi_1$, $\phi_2$, and $\phi_3$ at each arm.}
   \label{fig1}                                                                                                          
       \end{figure}

Recent advancements in quantum-dot systems offer a promising platform for investigating the 
fusion and braiding of Majorana zero modes (MZMs). These processes are important for the testing of 
non-Abelian statistics of MZMs. Successful quantum computation with MZMs requires their adiabatic manipulation. 
Quantum-dot systems (see Fig.~\ref{fig1}) enable such experiments even in small-scale setups, as MZMs are fully 
localized at individual sites, unlike in nanowire systems~\cite{Dvir,ChunX}.
In these quantum-dot systems, the on-site chemical potential can be finely tuned using quantum gates, 
facilitating precise control over the adiabatic movement of MZMs. 
However, performing braiding experiments in strictly one-dimensional geometries is challenging, 
as MZMs can fuse during their exchange~\cite{Alicea}. To conduct braiding experiments effectively, $Y$- or $X$-shaped 
geometries are required~\cite{Tong,Bpandey}.
Moreover, the realization of MZMs at the sweet-spot ($t_h = \Delta $) also pave the way  
to perform analytical calculations for solving the MZM wavefunctions in these complex geometries~\cite{Bpandey,pandey3}. 
Understanding the Majorana wavefunctions near the junctions of $Y$- or $X$-shaped configurations 
is important for the successful fusion and braiding of MZMs.

The organization of the manuscript is as follows. 
First, we will discuss the one-dimensional quantum dot system with the same $p$-wave pairing phases. 
Then, we will describe the exact form of Majorana wavefunctions near the junction of $Y$-shaped geometry quantum dots. 
Next, we will address the time-dependent trivial fusion of MZMs in a one-dimensional geometry using a time-dependent moving wall. 
Finally, we will present the non-trivial fusion of MZMs using two pairs of MZMs, where the two pairs have predefined parities.

\section{Results}
\subsection{One-dimensional Kitaev chain in quantum-dot systems}
The recent advancements in quantum-dot systems lead to the 
experimental realization of two- and three-sites Kitaev chains~\cite{Dvir,ChunX}. 
In these systems, the spin-polarized quantum-dots are coupled through superconductor-semiconductor hybrids 
and are well controlled by electrostatics gates~\cite{Dvir,ChunX}. This setup  allowed the single electron hopping $t$ (ECT) and the
triplet pairing  $\Delta$ to be controlled  (by the subgap Andreev bound states residing in the hybrid segments~\cite{Liu})  
providing all the required elements to realize artificial Kitaev chains~\cite{Dvir}.
The Kitaev chain Hamiltonian for spinless fermions with hopping $t$ and $p$-wave pairing $\Delta$ can be described as~\cite{Kitaev1}:
    \begin{equation} 
      H= \sum_j^{N-1} \left(-tc^{\dagger}_j c^{\phantom{\dagger}}_{j      +1} +|\Delta|e^{i\phi_j}c_jc_{j+1} +h.c \right),
     \end{equation}
   where $c_j$ is a spinless fermionic annihilation operator at site $j$ and $\phi_k$ is the pairing phase on bond $k$.
   Using the relation $c_j= \frac{1}{\sqrt{2}} e^{-i\phi_k/2}\left(\gamma^I_{A,j}+i\gamma^{I}_{B,j}\right)$, 
   at the sweet-spot ($t=\Delta$) and at $\phi_k=0$ the Kitaev chain Hamiltonian in terms of Majorana operators can be written as:
     \begin{equation}    
       H=-2i\Delta \sum_j^{N-1}\gamma^I_{A,j+1}\gamma^I_{B,j}.
     \end{equation}
     Interestingly, the Majorana operators $\gamma^I_{A,1}$ and $\gamma^I_{B,N}$
     are absent in the Hamiltonian and satisfy $\gamma^{I \dagger}_{A,1}= \gamma^I_{A,1}$, and $\gamma^{I \dagger}_{B,N}= \gamma^I_{B,N}$. 
     Thus, these two end Majorana operators  $\gamma^I_{A,1}$ and $\gamma^I_{B,N}$  represent Majorana zero modes, as intuitively expected.
      The signature of Majorana zero modes has been confirmed  
      via tunneling conductance experiments at the sweet-spot ($t=\Delta$) in the quantum-dot experiments~\cite{Dvir,ChunX}.

\subsection{$Y$-shape Kitaev wire in quantum-dot system}

The $Y$-shaped geometry plays a crucial role in facilitating the potential 
braiding of Majorana Zero Modes (MZMs) in quantum-dot experiments~\cite{Bpandey}. 
We analytically solve the $Y$-shaped Kitaev wires at the sweet spot, 
considering various superconducting phases in each arm. 
Remarkably, by expressing the Majorana operators as 
\( c_j = \frac{1}{\sqrt{2}} e^{-i\phi_k/2} \left( \gamma^I_{A,j} + i\gamma^{I}_{B,j} \right) \), 
we can write four distinct independent Hamiltonian's:
one for each of the three arms (I, II, III) and one for the central region (IV). 
This approach allows us to independently diagonalize and precisely solve each region at the sweet-spot~\cite{Bpandey}. 
Depending on the superconducting phases in each arm, we observe the emergence of exotic multi-site Majorana zero modes near the central region.

For the phases \(\phi_1 = \pi\), \(\phi_2 = 0\), and \(\phi_3 = 0\), 
we identify a total of six MZMs. 
Specifically: (a) three single-site MZMs are located at the edge sites, (b) one single-site MZM is found at the central site,
and (c) two multi-site MZMs are present near the central region. 
In contrast, when the phases are equal (\(\phi_1 = 0\), \(\phi_2 = 0\), and \(\phi_3 = 0\)), 
we observe a total of four MZMs: (a) similar to the previous scenario, three single-site MZMs are localized at the edge sites, 
but (b) only one multi-site MZM is found near the central region.
We further analyzed the stability of single- and multi-site MZMs against the repulsive interaction $(V)$
by separately calculating the electron and hole components of $(LDOS(\omega, j))$ using by the density-matrix renormalization group method (DMRG).
For moderate values of repulsive interaction $(V)$, our DMRG results show that 
the single-site edge MZMs and multi-site MZMs exhibit nearly equivalent stability~\cite{Bpandey}.

In the context of braiding experiments, where it is crucial that Majorana wave functions do not overlap,
it is essential to understand the shape of multi-site Majorana wave functions when exchanging MZMs near the junction. 
Our findings regarding the existence and stability of multi-site MZMs, especially against Coulomb repulsion and deviations
from the sweet spot, are valuable for developing fully functional $Y$-shaped junctions composed of quantum-dot arrays.
These multi-site MZMs should be detectable in quantum-dot experiments near the sweet spots,
utilizing just seven quantum dots arranged in a $Y$-shaped geometry (Fig.~\ref{fig1}(b)) for tunneling-conductance measurements.

\subsection{Time-dependent fusion of MZMs in quantum-dots}

\begin{figure} [ht]                                                                                                            
     \begin{center}                                                                                                                 
    \begin{tabular}{c} 
      \includegraphics[height=5cm]{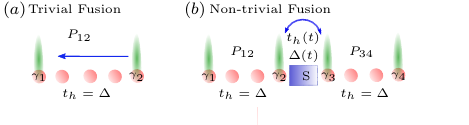}                                                                                        
      \end{tabular}                                                                                                                  
      \end{center}                                                                                                                   
\caption{(a) Sketch of the trivial fusion using two Majorana zero modes at the sweet-spot $t_h=\Delta$.
    (b)Sketch of the non-trivial fusion using two separate pairs of MZMs at the sweet-spot.
    The left part of the system has  a pre-defined parity $P_{12}$ while the right part has parity $P_{34}$.}
    \label{fig2}                                                                                                          
       \end{figure}

Recent advancements in quantum-dot systems have created a promising platform for testing the non-Abelian statistics of MZMs~\cite{Dvir,ChunX}. In these systems, MZMs can be fully localized on a single site at the sweet spot, 
enabling the study of their fusion and braiding even in small setups. 
Fusion and detection of MZMs in quantum-dot systems are more convenient compared to braiding experiments, 
which require complex geometries. MZMs follow the fusion rule ($\gamma$ $\times$ $\gamma$ = $I$ + $\Psi)$,
meaning two MZMs ($\gamma$) can result in either a vacuum state ($I$)
or a fermion ($\Psi$), demonstrating their non-Abelian nature~\cite{Aasen}.
The quantum-dot systems allow for two distinct fusion methods: $``$trivial" and $``$non-trivial" 
(Fig.~\ref{fig2}). Understanding these methods  will be useful for the readout of the topological quantum 
states in the topological quantum computing
processes.

\subsubsection{Trivial fusion of MZMs in quantum-dots}

To perform a trivial fusion of MZMs, 
one can utilize a single pair in a chain by moving one edge MZM 
towards the other using a moving potential wall (or well), as illustrated in Fig.~\ref{fig2}(a). 
This fusion process is deterministic as it occurs within the same pair with well-defined parity ($P_{12} = +1$ or $-1$). 
Following the trivial fusion, the MZMs either form a complete electron or a complete hole.
Achieving this fusion requires the slow movement of Majoranas, 
which can be controlled via time-dependent local gates connected to individual quantum dots. 
To monitor the dynamics and outcomes of this fusion, we calculated the time-dependent real-space 
local density of states, $LDOS(\omega,t,j)$, for both electrons and holes~\cite{Pandey}. 
By employing a time-dependent moving wall, as described in Ref.~\cite{Pandey}, 
we can move the right edge MZM to the left to facilitate their fusion. 
The fusion outcomes  depends on the initial state's parity~\cite{Pandey}.
For $P_{12} = -1$, the electron component of the local density-of-states exhibits a sharp peak near $\omega = 0$, 
and the charge density at site $j=1$ reaches one, indicating the formation of a single electron post-fusion.
Conversely, for $P_{12} = +1$, the charge density at site $j=1$ drops to zero, 
and the hole component of the local density-of-states shows a sharp peak near $\omega = 0$, signifying the formation of a hole.
To ensure the formation of a full electron post-fusion, 
we also determine the optimal speed range for moving the Majoranas. 
This calculation helps minimize nonadiabatic effects and prevent decoherence-induced poisoning (see Ref.~\cite{Pandey}).

\subsubsection{Non- trivial fusion of MZMs in quantum-dots}
To perform  the non-trivial fusion one needs to fuse the  Majoranas belonging to different  
     pairs of MZMs each with pre-defined parities. Interestingly, in the non-trivial case,                               the fusion outcomes are $probabilistic$, not deterministic~\cite{Aasen,Tsintzis,Zhou}. Compared to the trivial case, the non-trivial fusion
    can produce both electron and hole.  As shown in Fig.~\ref{fig2}(b), to perform non-trivial fusion of MZMs,
    we need at least  two-pairs of MZMs at the sweet-spot $t_h=\Delta$ (four MZMs total).  
We probe the fusion outcomes
    by calculating the time-dependent local density-of-states focusing on the central sites, using
     models simulating interacting quantum-dot systems. In the quantum-dot experiments, the hopping and 
     superconducting coupling between the quantum dots can be tuned by changing the electrostatic gates~\cite{Liu}.                                                           To observe the time-dependent non-trivial fusion,                                                                   
    we tune the time-dependent hopping and superconducting terms for the central part of the quantum-dot arrays, 
     (see Fig.~\ref{fig2}(b)). Using the time-dependent exact-diagonalization method  for the interacting electrons, 
     we  studied the non-trivial fusion of MZMs~\cite{pandey2}. In the case of the two pairs of MZMs initialized 
     with the same pairing phases $\phi_1=\phi_2=0$,  with increase in time, we find equal height peaks in the electron
     and hole components of the LDOS($\omega,t$). These results show the formation of 
     both electron $\Psi$ and vacuum channels  during the non-trivial fusion of MZMs. 
      We also have studied the Majorana fusion in a $\pi$-junction setup, 
      where we consider the pairing term $-\Delta$ for the left array, 
      and $+\Delta$ for the right array. With increase in the time-dependent hopping and pairing terms $\Delta(t)$
      between the two initialized pairs of MZMs, we find  that the MZMs near the $\pi$-junction do not fuse.
     Surprisingly, one MZM remains a localized single-site MZM, and another transforms into a multi-site MZM.
     The multi-site MZM is located on two sites with equal amplitude~\cite{pandey2}. 

     \section{Conclusion}

Quantum-dot systems provide an adaptable  platform for investigating the dynamics, fusion, and braiding of Majorana zero modes (MZMs). 
At the sweet-spot where ($t_h = \Delta$), we obtain an exact solution for a 
$Y$-shaped quantum-dot configuration. This study marks the first observation of {\it multi-site} Majorana zero modes within
a $Y$-shaped quantum wire. Our findings suggest that both single-site edge MZMs and multi-site MZMs exhibit similar 
stability against Coulomb repulsion. Through time-dependent local density-of-states analysis, we track the movement and fusion of Majoranas. 
In cases of trivial fusion, determined by the initial state parity, the outcome is either a complete electron or a complete hole. 
Conversely, non-trivial fusion results in a probabilistic outcome, with equal likelihoods of forming electrons or holes. 
The recent experimental progress in quantum-dot systems paves the way for future braiding experiments 
using a minimal number of quantum dots arranged in a $Y$-shaped geometry.

\section*{ACKNOWLEDGMENTS}
We thank N. Mohanta, N. Kaushal, G. Alvarez, R.-X Zhang, and S. Okamoto for helpful discussion.
The work of B.P., and E.D. was supported by the U.S. Department of Energy, Office of Science, Basic Energy Sciences,
 Materials Sciences and Engineering Division. 



\begin{thebibliography}{99}
\bibitem{Kitaev1}Kitaev AY. {Unpaired Majorana fermions in quantum wires.}
\href{https://doi.org/10.1070/1063-7869/44/10S/S29}{\it Phys.-Usp.} {\textbf{44}, 131 (2001)}.

\bibitem{Kitaev2}Kitaev AY. {Fault-tolerant quantum computation by anyons.}
\href{https://doi.org/10.1016/S0003-4916(02)00018-0}{\it Ann Phys (NY)} {\textbf{303}, 2 (2003)}.

\bibitem{brien}O$^,$Brien, T. E., Ro\.{z}ek, P., Akhmerov, A. R. {Majorana-Based Fermionic Quantum Computation.}
  \href{https://doi.org/10.1103/PhysRevLett.120.220504} {\it Phys. Rev. Lett.} {\textbf{120}, 220504 (2018)}.

\bibitem{Sarma}Sarma, S., Freedman, M.  Nayak, C. {Majorana zero modes and topological quantum computation.}
\href{https://doi.org/10.1038/npjqi.2015.1}{\it  npj Quantum Inf } {\textbf{1}, 15001 (2015)}.


\bibitem{Nayak}Nayak C, Simon SH, Stern A, Freedman M,  Sarma SD. {Non-abelian anyons and topological quantum computation.}
\href{https://doi.org/10.1103/RevModPhys.80.1083}{\it Rev Mod Phys } {\textbf{80}, 1083 (2008)}.


\bibitem{Roman} Lutchyn, R. M., Sau, J. D., and Sarma SD. {Majorana Fermions and a Topological Phase Transition in Semiconductor-Superconductor Heterostructures} \href{https://doi.org/10.1103/PhysRevLett.105.077001} {\it Phys. Rev. Lett.} {\textbf{105}, 077001 (2010)}.

\bibitem{Mourik} Mourik, V. et al. {Signatures of Majorana fermions in hybrid superconductor-semiconductor nanowire devices.}  \href{https://doi.org/10.1126/science.1222360} {\it Science} {\textbf{336}, 1003–1007 (2012)}.

\bibitem{Stanescu}Sarma, SD., Sau, J. D., and Stanescu, T. D. {Spectral properties, topological patches, and effective phase diagrams of finite disordered Majorana nanowires.}  \href{https://doi.org/10.1103/PhysRevB.108.085416} {\it Phys. Rev. B} {\textbf{108}, 085416 (2023)}.

\bibitem{Gomanko}Yu, P., Chen, J., Gomanko, M. et al. {Non-Majorana states yield nearly quantized conductance in proximatized nanowires.} \href{https://doi.org/10.1038/s41567-020-01107-w}{\it Nat. Phys.} {\textbf{17}, 482–488 (2021)}.

\bibitem{Loo} van Loo, N., Mazur, G.P., Dvir, T. et al. {Electrostatic control of the proximity effect in the bulk of semiconductor-superconductor hybrids.} \href{https://doi.org/10.1038/s41467-023-39044-w} {\it Nat Commun.}  {\textbf{14}, 3325 (2023)}. 

\bibitem{PanH}Sarma, SD. and Pan, H. {Disorder-induced zero-bias peaks in Majorana nanowires.}  \href{https://doi.org/10.1103/PhysRevB.103.195158} {\it Phys. Rev. B} {\textbf{103}, 195158 (2021)}.



\bibitem{Jay}Sau, J., Sarma, S. {Realizing a robust practical Majorana chain in a quantum-dot-superconductor linear array.}
\href{https://doi.org/10.1038/ncomms1966}{\it  Nat Commun.} {\textbf{3}, 964 (2012)}.

\bibitem{Loss}Ran\v{c}i\'{c}, J. M., Hoffman, S., Schrade, C., Klinovaja, J., \&  Loss, D. 
{Entangling spins in double quantum dots and Majorana bound states.}
\href{https://doi.org/10.1103/PhysRevB.99.165306} {\it Phys. Rev. B.} {\textbf{99} 165306 (2019)}.

\bibitem{Deng} Deng, M. T., Vaitiek{\.{e}}nas, E., Hansen, E. B., Danon, J.  et al. {Majorana bound state in a coupled quantum-dot hybrid-nanowire system.}
\href{https://doi.org/10.1126/science.aaf3961} {\it Science} {\textbf{354}, 1557–1562 (2016)}.

\bibitem{Dvir}Dvir, T., Wang, G., van Loo, N. et al. {Realization of a minimal Kitaev chain in coupled quantum dots.}
\href{https://doi.org/10.1038/s41586-022-05585-1}{\it  Nature } {\textbf{614}, 445–450 (2023)}.

\bibitem{Bordin} Bordin, A., Li, X., Driel, D. V., Wolff, J. C.
 et al. {Crossed Andreev reflection and elastic co-tunneling in a three-site Kitaev chain nanowire device.}
\href{https://doi.org/10.48550/arXiv.2306.07696}{\it  arXiv:} {\textbf{2306}, 07696 (2023)}.

\bibitem{ChunX} Bordin, A.,  Liu, Chun-Xiao, Dvir, T., et al. {Signatures of Majorana protection in a three-site Kitaev chain}. 
\href{https://doi.org/10.48550/arXiv.2402.19382}{\it arXiv: } {\textbf{2402}, 19382 (2024}. 

\bibitem{Mazur}Wang, G., Dvir, T., Mazur, G.P. et al. {Singlet and triplet Cooper pair splitting in hybrid superconducting nanowires.} \href{https://doi.org/10.1038/s41586-022-05352-2} {\it Nature} {\textbf{612}, 448–453 (2022)}. 

\bibitem{Liu} Liu, C.-X., Wang, G., Dvir, T. \& Wimmer, M. 
{Tunable superconducting coupling of quantum dots via Andreev bound states in semiconductor-superconductor nanowires.}
\href{https://doi.org/10.1103/PhysRevLett.129.267701} {\it Phys. Rev. Lett.} {\textbf{129}, 267701 (2022)}

\bibitem{Alicea} Alicea, J., Oreg, Y., Refael, G. et al. {Non-Abelian statistics and topological quantum information processing in 1D wire networks.}
\href{https://doi.org/10.1038/nphys1915}{\it   Nature Phys.} {\textbf{7}, 412–417 (2011)}.

\bibitem{Tong}Zhou, T., Dartiailh, M. C.,  Mayer, W., Han, J. E., Matos-Abiague,  et al. 
{Phase Control of Majorana Bound States in a Topological X Junction}  
\href{https://doi.org/10.1103/PhysRevLett.124.137001}{\it Phys. Rev. Lett.} {\textbf{124}, 137001 (2020)}.

\bibitem{Bpandey} Pandey, B., Kaushal, N., Alvarez, G. et al. {Majorana zero modes in Y-shape interacting Kitaev wires.} 
  \href{https://doi.org/10.1038/s41535-023-00584-5} {\it npj Quantum Mater.} {\textbf{8}, 51 (2023).}

\bibitem{pandey3} B. Pandey, G. Alvarez, E. Dagotto, R.-X. Zhang,                                                                 
 \href{https://doi.org/10.48550/arXiv.2407.00158}{\it arXiv:} {\textbf{2407}, 00158 (2024)}.


\bibitem{Pandey} Pandey,B.,  Mohanta, N.,  Dagotto, E.{
Out-of-equilibrium Majorana zero modes in interacting Kitaev chains.}
\href{https://doi.org/10.1103/PhysRevB.107.L060304}{\it Phys. Rev. B.} {\textbf{107}, L060304 (2023)}.


\bibitem{Aasen} Aasen, D., Hell, M., Mishmash, R. V., Higginbotham, A., et al. {
Milestones toward Majorana-based quantum computing.}
\href{https://doi.org/10.1103/PhysRevX.6.031016}{\it Phys. Rev. X.} {\textbf{6}, 031016 (2016)}.

\bibitem{Tsintzis} Tsintzis, A., Souto,R. S.,  Flensberg, K., et al. {Majorana Qubits and Non-Abelian Physics in Quantum Dot–Based Minimal Kitaev Chains.} \href{https://doi.org/10.1103/PRXQuantum.5.010323}{\it PRX Quantum } {\textbf{5},  010323 (2024)}. 

\bibitem{Zhou} Zhou, T., Dartiailh, M.C., Sardashti, K. et al. {Fusion of Majorana bound states with mini-gate control in two-dimensional systems.}
  \href{https://doi.org/10.1038/s41467-022-29463-6} {\it Nat Commun} {\textbf{13}, 1738 (2022). }

\bibitem{pandey2} Pandey, B., Okamoto, S., and Elbio Dagotto {Unexpected results for the non-trivial fusion of Majorana zero modes in interacting quantum-dot arrays.}       \href{https://doi.org/10.48550/arXiv.2311.15079}{\it arXiv:} {\textbf{2311}, 15079 (2023).}


\end{thebibliography}


\end{document}